\documentclass[fleqn,10pt]{wlscirep}
\usepackage[utf8]{inputenc}
\usepackage{siunitx}
\usepackage{xcolor}
\usepackage[T1]{fontenc}
\title{Interplay of Umklapp scattering and Sb--Au hybridization in surface-reconstructed Sb/Au(111)}

\author[1,+]{Zhe Zheng}
\author[1,+]{Celine Wassenberg}
\author[1]{Stefanie Hilgers}
\author[1]{Carsten Westphal}
\author[1,*]{Mirko Cinchetti}
\affil[1]{Department of Physics, TU Dortmund University, 44227, Dortmund, Germany}

\affil[*]{mirko.cinchetti@tu-dortmund.de}

\affil[+]{these authors contributed equally to this work}

\begin{abstract}
Surface reconstructions induced by atomic adsorption can strongly reshape metallic surface states, providing a direct pathway to tune their electronic structure. Using angle-resolved photoemission spectroscopy, we investigate the electronic structure of Sb/Au(111) during the coverage-driven evolution from the clean Au(111) surface to the $(14\times14)$ and Rec$(3\times\sqrt{3})$ phases. In the Rec$(3\times\sqrt{3})$ phase, triangular Fermi pockets emerge at the Brillouin-zone boundary. Their momentum positions are consistent with a reciprocal-space folding construction, but their reduced size near the Fermi level indicates a modification of the Au-derived $sp$ dispersion. The substantial modifications of deeper Au $d$-derived bands observed in ARPES further indicate significant mixing between Sb $p$ orbitals and Au $d$ states. These results show that the electronic structure of Sb/Au(111) is governed by the interplay between reconstruction-induced Umklapp scattering and interfacial orbital hybridization, highlighting adsorbate–substrate hybridization as a key mechanism for tuning and engineering surface electronic structures.
\end{abstract}

\begin{document}    

\flushbottom
\maketitle
\thispagestyle{empty}

\section*{Introduction}

Surface reconstructions\cite{titmuss1996reconstruction,yang2019atomically} induced by atomic adsorption provide an effective route to engineering electronic structures at metal interfaces\cite{chen1998scanning,zhuang2018band}. The superlattice periodicities introduced by reconstructed surfaces can generate replica bands and folded dispersions in angle-resolved photoemission spectroscopy (ARPES) measurements\cite{sohail2015umklapp,jauernik2018probing,chiniwar2019substrate}, while interfacial interactions\cite{marchenko2012giant,ast2007giant} may simultaneously modify the underlying electronic states through orbital hybridization\cite{sakamoto2019impact,patil2016arpes,hu2023chiral,cantero2021synthesis,castro2025three} and many-body effects\cite{marsusi2018graphene,katoch2018giant,garcia2011renormalization}. Distinguishing geometric band folding from intrinsic electronic effects thus remains a central challenge in interpreting ARPES spectra of reconstructed surfaces\cite{watson2019probing,brouet2025unfolding,chiniwar2019substrate}.

Two-dimensional materials composed of group-VA elements have attracted significant attention due to their tunable electronic structures\cite{zhou2019interfacial,khan2021novel,zhang2017topologically,lu2022realization}, strong spin--orbit coupling\cite{lee2015two}, and structural anisotropy\cite{lei2019anisotropic,zhu2020kagome}. Following the discovery of graphene\cite{novoselov2004electric}, research on elemental two-dimensional materials has expanded from group-IV analogues such as stanene\cite{zhu2015epitaxial,maniraj2019case} to group-VA systems including phosphorene\cite{zhuang2018band}, antimonene\cite{ji2016two,lu2021observation}, and bismuthene\cite{lu2024realization}. Epitaxial growth on metal substrates is a common route to stabilize such atomic layers and to tune their interface-dependent properties\cite{zhu2015epitaxial,maniraj2019case,zhou2019interfacial,cantero2021synthesis}. In particular, Au(111) provides a well-established platform for investigating adsorption-driven surface reconstructions and the emergence of interface-derived electronic states in low-dimensional materials. Previous studies have shown that Sb adsorption on Au(111) depends strongly on coverage and annealing conditions\cite{cantero2021synthesis,zan2019antimony,ma1993adsorption}, leading to intermediate phases, long-period reconstructions, and Au$_2$Sb surface-alloy motifs\cite{hu2025unconventional}. This competition between alloy formation and two-dimensional layer growth is also reported for Sn/Au(111), where Au$_2$Sn alloy formation and subsequent two-dimensional Sn growth are strongly affected by the Au(111) reconstruction\cite{shah2021atomic,hochhaus2025square,hochhaus2025structural}. These structural variations lead to complex surface electronic structures in which substrate states, reconstruction-induced replicas, and adsorbate-derived bands may coexist\cite{castro2025three,pierron2026surface}.

Despite these advances, the electronic structure of Sb/Au(111) remains incompletely understood. Previous ARPES studies have revealed Sb-derived states\cite{hu2023chiral} as well as reconstruction-driven replica features in $\alpha$-antimonene/Au(111)\cite{pierron2026surface}. Importantly, these studies do not require a purely Umklapp-based interpretation; instead, they emphasize that surface-reconstruction-driven band folding and strong Sb--Au hybridization can both contribute to the measured spectra\cite{pierron2026surface,castro2025three}. In reconstructed Sb overlayers, multiple rotational domains and coexisting superlattice periodicities further complicate the assignment of spectral features. A systematic understanding of how geometric band folding, hybridization, and band renormalization evolve across the coverage-driven structural sequence of Sb/Au(111) is therefore still lacking.

In this work, we investigate Sb/Au(111) across the coverage-dependent structural evolution from the clean Au(111) surface to the $(14\times14)$ and Rec$(3\times\sqrt{3})$ phases. By combining ARPES measurements with structural characterization by Low-energy electron diffraction (LEED), we analyze the origin of the electronic features emerging in the reconstructed phases. Our results reveal triangular Fermi pockets associated with the superlattice potential, while their evolution near the Fermi level indicates a modification of the Au-derived $sp$ dispersion beyond a purely geometric folding model. At deeper binding energies, significant changes of Au-derived valence states provide further evidence of orbital hybridization between Sb $p$ orbitals and Au $d$ orbitals. These findings demonstrate that the electronic structure of Sb/Au(111) is governed by the interplay between reconstruction-induced Umklapp scattering and Sb--Au hybridization, providing a promising platform for band engineering and Fermi-surface topology control in low-dimensional materials.

\section*{Results}

\subsection*{Sample growth and structural characterization}

\begin{figure}[ht]
\centering
\includegraphics[width=0.95\linewidth]{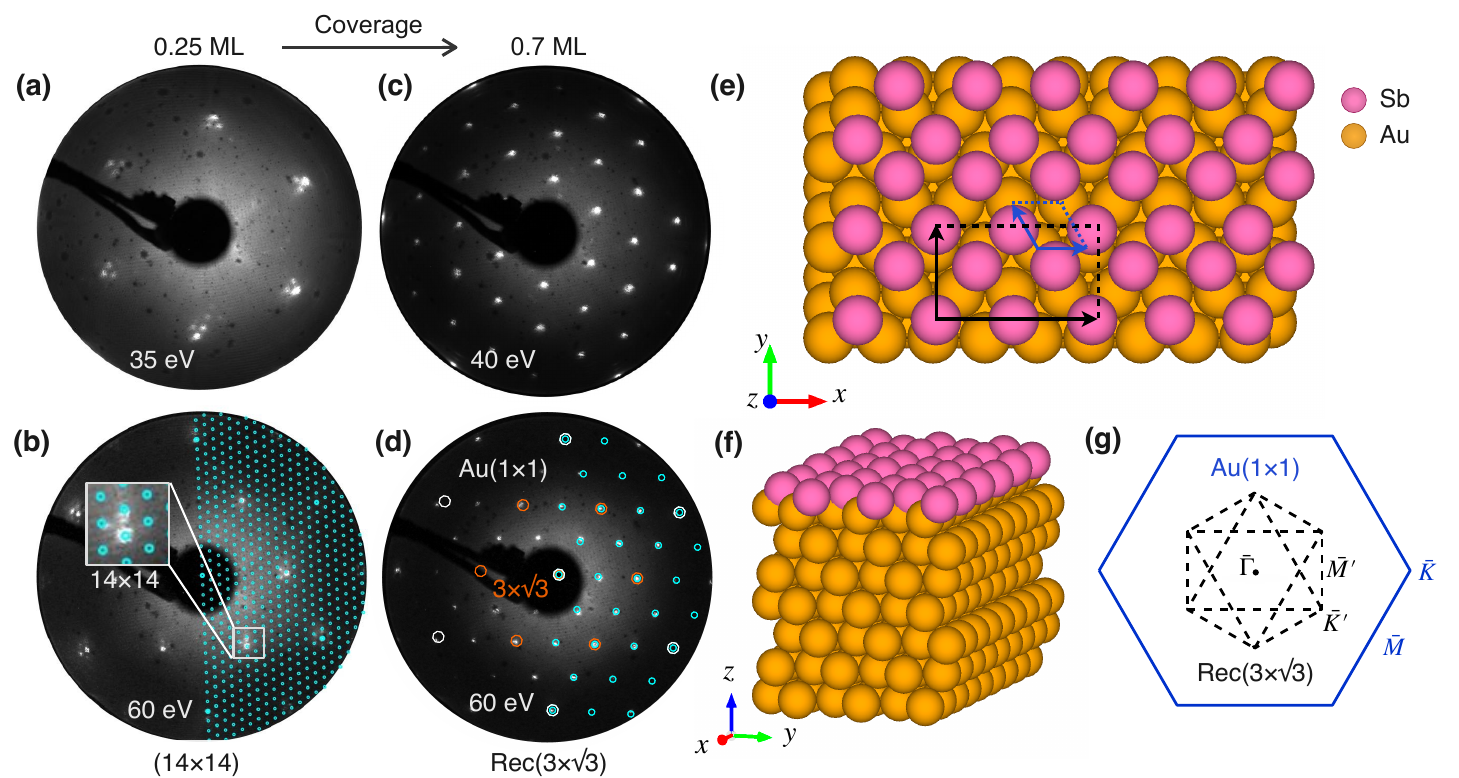}
\caption{\textbf{Structural evolution of Sb on Au(111).}{(a–d) LEED patterns of Sb/Au(111) at different Sb coverages. 
At 0.25 ML, a long-period $(14\times14)$ reconstruction is observed as satellite spots around the Au(1$\times$1) reflections. With increasing coverage (0.7 ML), the pattern evolves into a well-defined Rec$(3\times\sqrt{3})$ superstructure, indicating the formation of an ordered Sb overlayer. (b) and (d) show representative LEED patterns at 60 eV with overlays marking the reciprocal lattice vectors of the $(14\times14)$ reconstruction and the Rec$(3\times\sqrt{3})$ phase, respectively. (e,f) Structural model of Sb/Au(111) in the Rec$(3\times\sqrt{3})$ phase, adapted from the Sb overlayer structure reported by Cantero \textit{et al.}\cite{cantero2021synthesis}. Sb atoms (pink) form an ordered overlayer on Au(111) (orange). (e) Top view; (f) side view. Arrows in (e) mark the primitive unit cells of Au (blue) and Sb (black). (g) Surface Brillouin zones of Au(1$\times$1) and the Sb$(3\times\sqrt{3})$ superstructure with three $120^\circ$ rotational domains.}}
\label{fig:Fig1}
\end{figure}

{To establish the structural evolution of Sb on Au(111), Sb was deposited onto Au(111) from a Knudsen cell held at $420\,^\circ\mathrm{C}$ and the samples were annealed for $\SI{30}{\minute}$ at $260\,^\circ\mathrm{C}$. LEED patterns show that the final surface phase is solely determined by the coverage. At 0.25 ML, satellite spots appear around the Au(1$\times$1) reflections, indicating a long-period $(14\times14)$ reconstruction, as shown in Figure~\ref{fig:Fig1}(a,b). Increasing the coverage to 0.7 ML yields a well-defined Rec$(3\times\sqrt{3})$ superstructure as shown in Figure~\ref{fig:Fig1}(c,d). The reciprocal-space overlays in Figure~\ref{fig:Fig1}(b,d) establish the indexing of both phases.}

{The structural model in Figure~\ref{fig:Fig1}(e,f) represents the Rec$(3\times\sqrt{3})$ phase as an ordered Sb overlayer on Au(111) and is adapted from the model proposed by Cantero \textit{et al.}\cite{cantero2021synthesis}. In the top view (Figure~\ref{fig:Fig1}(e)), the primitive unit cells of the Au substrate and the Sb overlayer are indicated, making the relation between the overlayer periodicity and the substrate lattice explicit. The surface Brillouin-zone construction in Figure~\ref{fig:Fig1}(g) shows that the Rec$(3\times\sqrt{3})$ overlayer generates a reduced surface Brillouin zone and forms three rotational domains related by $120^\circ$ symmetry. These domains should be considered when interpreting the ARPES spectra, as they superpose symmetry-related spectral weight in momentum space.}

\subsection*{Band folding and Umklapp scattering in reconstructed Sb/Au(111)}

\begin{figure}[ht]
\centering
\includegraphics[width=0.95\linewidth]{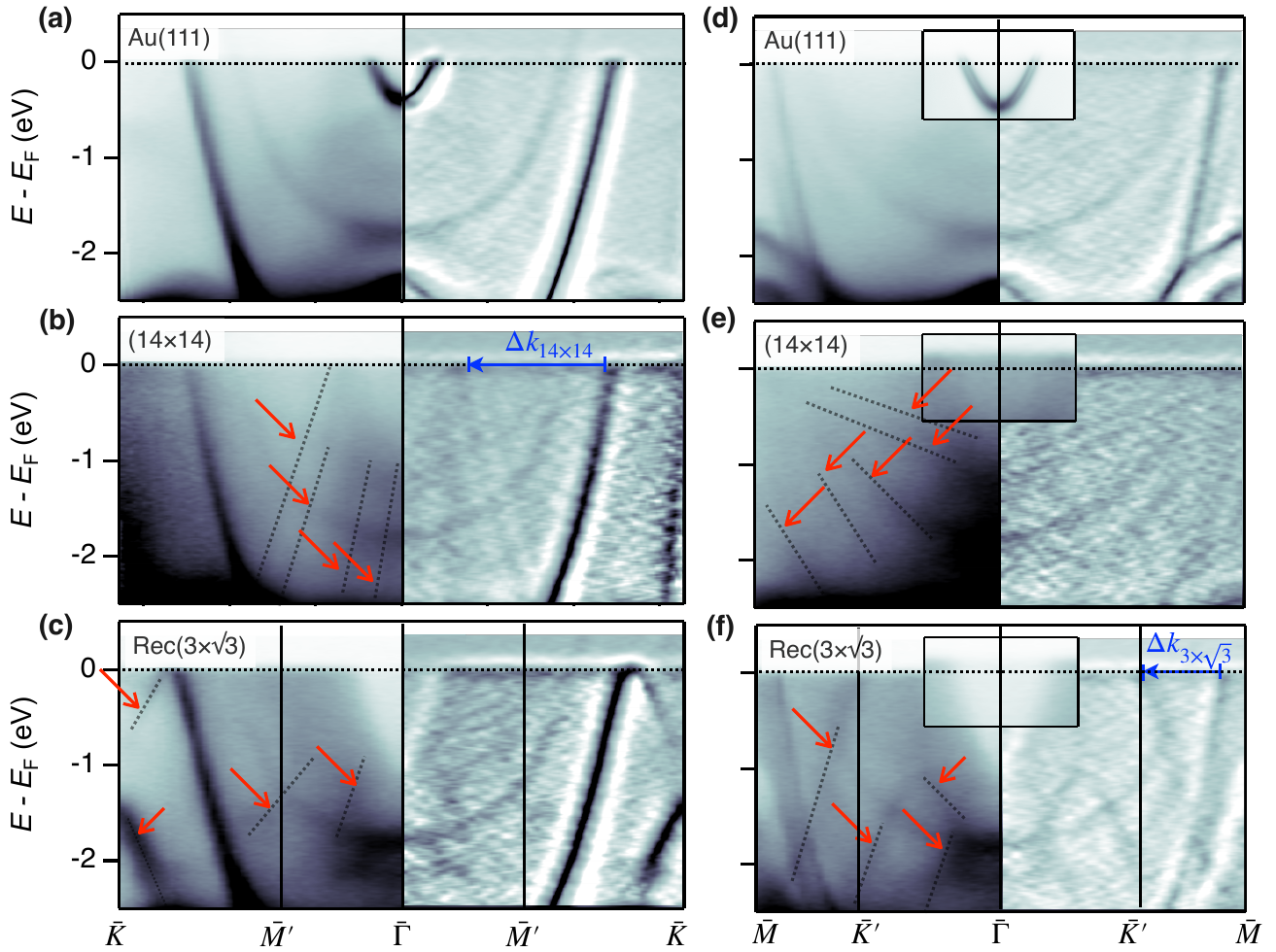}
\caption{\textbf{Coverage-dependent evolution of the electronic structure and band folding in Sb/Au(111).} (a--f) ARPES band dispersions measured along the $\bar{K}-\bar{M}'-\bar{\Gamma}$ and $\bar{M}-\bar{K}'-\bar{\Gamma}$ directions for (a,d) clean Au(111), (b,e) the $(14\times14)$ Sb/Au(111) phase, and (c,f) the Rec$(3\times\sqrt{3})$ Sb/Au(111) phase. In each panel, the left side shows the raw dispersion and the right side the corresponding curvature plot~\cite{zhang2011precise}. Insets highlight the strong suppression and disappearance of the surface state near $\bar{\Gamma}$ upon Sb adsorption. Upon formation of the $(14\times14)$ reconstruction, the superlattice potential folds the Au(111) bands, giving rise to additional dispersive features (red arrows). In the Rec$(3\times\sqrt{3})$ phase, these folded bands become more pronounced, indicating a stronger superlattice potential and enhanced Sb--Au hybridization. In panels (b) and (f), blue arrows mark representative Umklapp scattering processes of the Au-derived $sp$ band, described by $\mathbf{k}'=\mathbf{k}+p\mathbf{G}^{\mathrm{Sb}}+q\mathbf{G}^{\mathrm{Au}}_{1\times1}$. The corresponding momentum shifts, $\Delta \mathbf{k}_{14\times14}=-0.715\text{(1/\AA)}$ and $\Delta \mathbf{k}_{3\times\sqrt{3}}=-0.455\text{(1/\AA)}$, quantitatively reproduce the observed replica-band displacements, confirming their origin in superlattice-induced Umklapp scattering.}
\label{fig:Fig2}
\end{figure}

ARPES measurements resolve a systematic evolution of the Au(111) electronic structure with increasing Sb coverage, as shown in Fig.~\ref{fig:Fig2} and in Figs.~S1 and S4 of the Supplementary Information. For the pristine Au(111) surface, the spectra exhibit the characteristic Shockley surface state\cite{lashell1996spin,reinert2001direct,tusche2015spin} centered at $\bar{\Gamma}$ with a nearly parabolic dispersion, together with an additional Au-derived $sp$ band\cite{courths1986electronic,reinert2001direct,sheverdyaeva2016energy} at larger momenta, as shown in Figure~\ref{fig:Fig2}(a,d). Upon Sb adsorption and formation of the $(14\times14)$ reconstruction, additional dispersive features appear, as shown in Figure~\ref{fig:Fig2}(b,e). These features arise from the periodic potential of the superlattice, which induces band folding of the Au electronic states via Umklapp scattering processes described by $\mathbf{k}'=\mathbf{k}+p\mathbf{G}^{\mathrm{Sb}}+q\mathbf{G}^{\mathrm{Au}}_{1\times1}$, with $p,q\in\mathbb{Z}$. 
As a representative example, the momentum displacement of the folded branch in the $(14\times14)$ phase is approximately $-0.715\text{(1/\AA)}$, consistent with the separation between the original and folded Au-derived $sp$ bands observed in Figure~\ref{fig:Fig2}(b). With further Sb deposition, the system transitions to the Rec$(3\times\sqrt{3})$ reconstruction, where a different set of folded branches appears, as shown in Figure~\ref{fig:Fig2}(c,f). A characteristic displacement of approximately $-0.455\text{(1/\AA)}$ is reproduced by the reciprocal-space folding construction described in Supplementary Sec.~S2. In the folding simulation this displacement arises from translations of the Au-derived contour into repeated zones of the Sb superstructure. The observed replica positions are therefore consistent with superlattice-induced band folding, while their spectral weight and energy evolution require the additional interfacial effects discussed below.

\begin{figure}[ht]
\centering
\includegraphics[width=\linewidth]{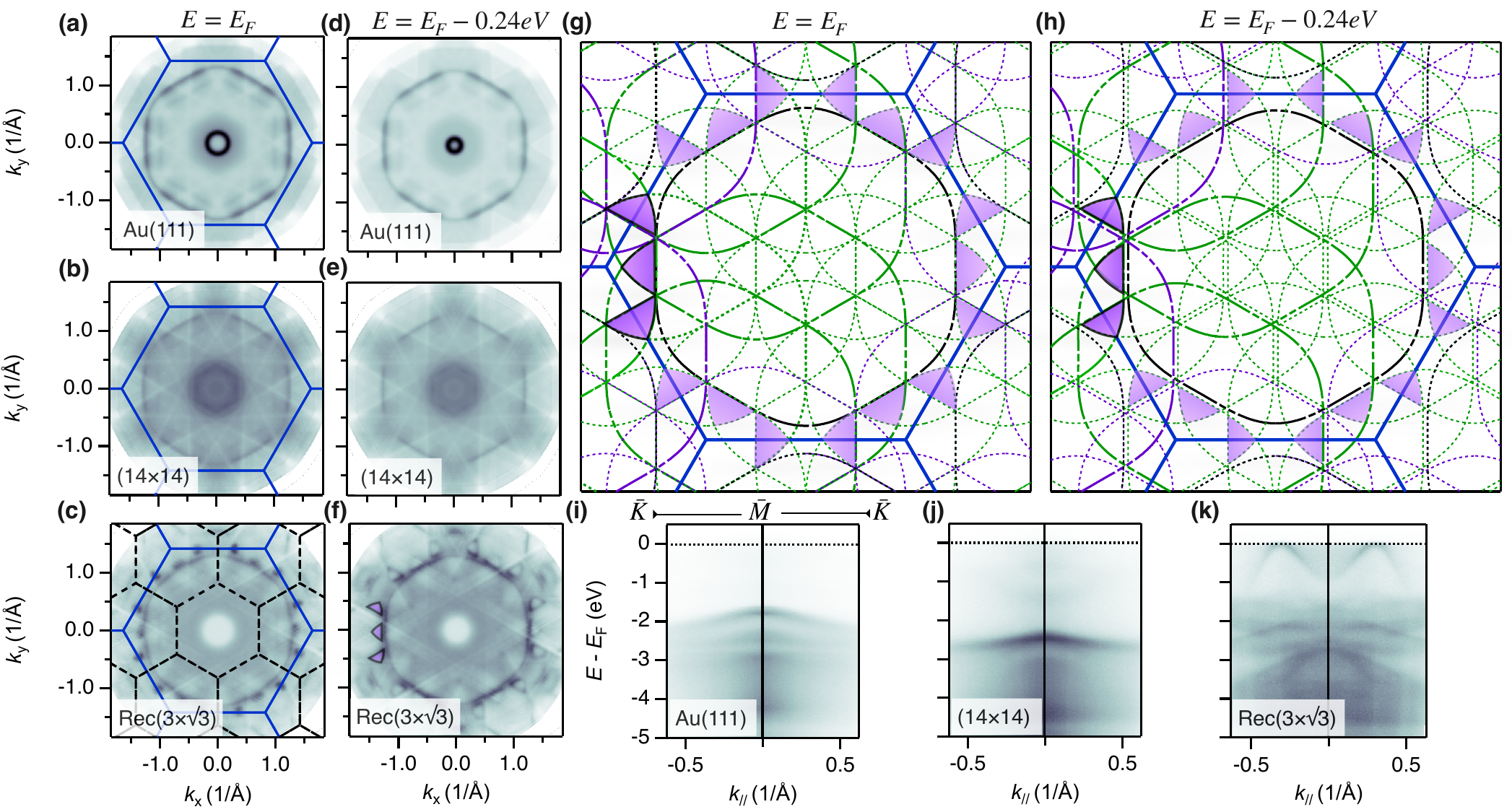}
\caption{\textbf{Constant-energy maps, band dispersions and simulated Umklapp scattering patterns of Sb/Au(111).} (a--f) ARPES constant-energy maps measured at $E=E_F$ and $E=E_F-0.24\,\mathrm{eV}$ for (a, d) clean Au(111), (b, e) the $(14\times14)$ Sb/Au(111) phase, (c, f) the Rec$(3\times\sqrt{3})$ Sb/Au(111) phase. The blue hexagon denotes the Au(111) surface Brillouin zone, while the black dashed hexagon marks the Rec$(3\times\sqrt{3})$ Sb surface Brillouin zone for the three equivalent rotational domains. Additional features emerging with increasing Sb coverage originate from superlattice-induced band folding and hybridization. (g,h) Simulated patterns obtained by applying Umklapp scattering of the Au $sp$ surface state by the reciprocal lattice vectors of the Rec$(3\times\sqrt{3})$ Sb superstructure with three rotational domains. The construction captures the positions of the triangular pockets, while the reduced pocket size at $E_F$ shows that geometric folding alone is insufficient. (i--k) ARPES band dispersions measured along the $\bar{K}-\bar{M}-\bar{K}$ direction for (i) Au(111), (j) the $(14\times14)$ Sb/Au(111), and (k) the Rec$(3\times\sqrt{3})$ Sb/Au(111), respectively.}
\label{fig:Fig3}
\end{figure}

The effect of the superstructure becomes particularly evident in the constant-energy maps shown in Figure~\ref{fig:Fig3}. Whereas the clean Au(111) surface exhibits a nearly circular Fermi contour with hexagonal warping, as shown in Figure~\ref{fig:Fig3}(a,d), the Rec$(3\times\sqrt{3})$ phase gives rise to multiple triangular electron pockets in momentum space, as shown in Figure~\ref{fig:Fig3}(c,f). These features can be captured by an Umklapp construction in which the Au $sp$ state is scattered by the reciprocal lattice vectors of the Rec$(3\times\sqrt{3})$ reconstruction with three rotational domains. Details of the Fermi-contour fitting and the reciprocal-space construction are provided in Supplementary Information.

The Umklapp-scattering construction reproduces the triangular pockets reasonably well at $E = E_F - 0.24$~eV (Fig.~\ref{fig:Fig3}(f,h)), but not at the Fermi level (Fig.~\ref{fig:Fig3}(c,g)). In a purely geometric folding picture, the larger Au-derived $sp$ contour at $E_F$ would generate larger triangular pockets (Fig.~\ref{fig:Fig3}(g)). Experimentally, however, these pockets shrink markedly as $E_F$ is approached (Fig.~\ref{fig:Fig3}(c)). This behavior follows naturally from the downward-opening Au-derived parabola, whose constant-energy contour expands with increasing binding energy and therefore produces larger folded pockets away from $E_F$. A similar discrepancy was recently reported for $\alpha$-antimonene/Au(111), where the trigonal pockets were attributed not to simple Umklapp replicas, but to Au-derived states modified by the Sb-induced surface reconstruction and Sb--Au hybridization~\cite{pierron2026surface}. Therefore, this discrepancy indicates
that the reconstructed electronic structure cannot be understood in terms of geometric band folding alone, but requires an additional modification of the Au-derived dispersion, consistent with Sb–Au hybridization or reconstruction-induced band
modification.

\subsection*{Sb–Au hybridization in the deep-energy electronic states}

\begin{figure}[ht]
\centering
\includegraphics[width=0.9\linewidth]{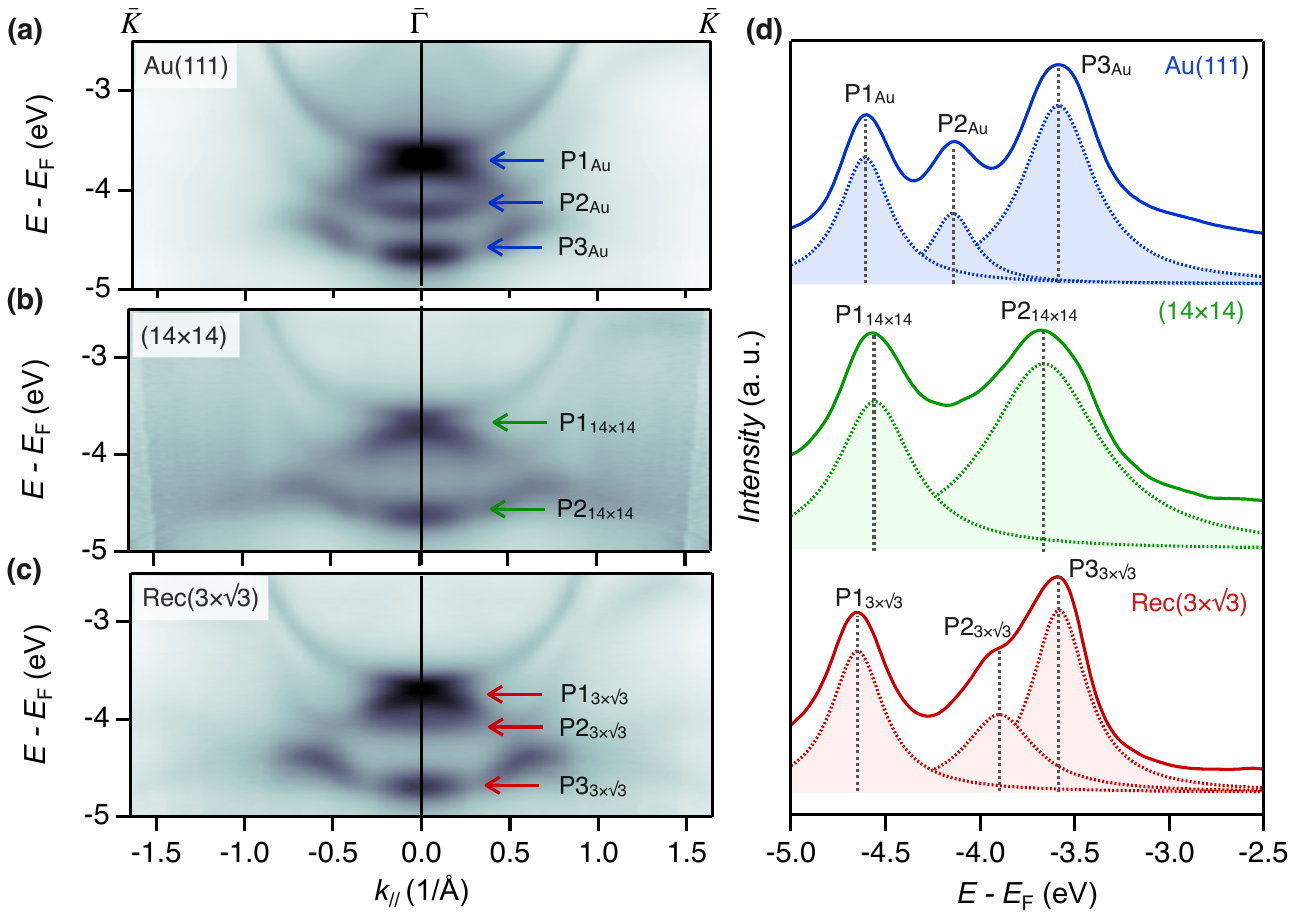}
\caption{\textbf{Evolution of deep valence states in Sb/Au(111).} (a–c) ARPES intensity maps along the $\bar{K}$–$\bar{\Gamma}$–$\bar{K}$ direction for (a) pristine Au(111), (b) the $(14\times14)$ Sb/Au(111) phase, and (c) the Rec$(3\times\sqrt{3})$ Sb/Au(111) phase. Several features located between $-5$ and $-2.5$ eV below $E_F$ are marked as P1–P4. (d) Energy distribution curves (EDCs) extracted near the $\bar{\Gamma}$ point comparing Au(111), $(14\times14)$, and Rec$(3\times\sqrt{3})$ phases. The peak positions (P1--P4) are determined from fits to the EDCs using Voigt components together with a Shirley-type background. The evolution of these peaks highlights systematic changes in the deep valence region upon Sb adsorption.}
\label{fig:Fig4}
\end{figure}

Insight into the influence of Sb adsorption on deeper valence states is provided by the ARPES intensity measured along $\bar{K}$–$\bar{\Gamma}$–$\bar{K}$ in the energy range from $-5$ to $-2.5$~eV relative to $E_F$, as shown in Figure~\ref{fig:Fig4}. For pristine Au(111), this region exhibits several broad features centered near $\bar{\Gamma}$, labeled P1$_{\text{Au}}$--P3$_{\text{Au}}$ in Figure~\ref{fig:Fig4}(a), which mainly originate from Au $5d$-derived states, as reported in previous photoemission measurements and band-structure calculations~\cite{sheverdyaeva2016energy,smith1974photoemission}. Upon formation of the $(14\times14)$ Sb/Au(111) phase, these features broaden and their spectral weights are redistributed, accompanied by a strong suppression at the binding energy around $-4.1$~eV. In the Rec$(3\times\sqrt{3})$ phase, the spectral line shape evolves further and an additional component (P2$_{3\times\sqrt{3}}$) emerges at an intermediate binding energy near $-3.9$~eV, as shown in Figure~\ref{fig:Fig4}(c). The energy distribution curves (EDCs) in Figure~\ref{fig:Fig4}(d) reveal pronounced peak shifts accompanied by substantial spectral-weight redistribution as the reconstruction develops. Such behavior is not expected from geometric band folding or Umklapp scattering alone\cite{anderson1976chemisorption,chiniwar2019substrate,pierron2026surface}, which primarily generate momentum replicas without modifying the intrinsic energy positions or spectral line shapes. The emergence of P2$_{3\times\sqrt{3}}$ in the Rec$(3\times\sqrt{3})$ phase therefore indicates a modification of the underlying electronic states.

To quantify the evolution of the deeper valence states upon Sb adsorption on Au(111), representative EDCs extracted around $\bar{\Gamma}$ were fitted using Voigt components with a Shirley-type background\cite{shirley1972high,damascelli2003angle,hufner2013photoelectron} (see Supplementary Information for details). For pristine Au(111), three components reproduce the spectral line shape, yielding peak positions at $E_1=-4.61$~eV (FWHM $=0.33$~eV), $E_2=-4.14$~eV (FWHM $=0.28$~eV), and $E_3=-3.59$~eV (FWHM $=0.46$~eV). In the $(14\times14)$ phase, the spectral weight redistributes and the EDC is dominated by two components located at $E_1=-4.56$~eV (FWHM $=0.48$~eV) and $E_2=-3.66$~eV (FWHM $=0.72$~eV). For the Rec$(3\times\sqrt{3})$ phase, three components are again resolved at $E_1=-4.65$~eV (FWHM $=0.34$~eV), $E_2=-3.90$~eV (FWHM $=0.45$~eV), and $E_3=-3.58$~eV (FWHM $=0.35$~eV). The values in parentheses denote the full width at half maximum (FWHM) of each fitted component. Compared with pristine Au(111), the spectral line shape in the reconstructed phases exhibits clear modifications, reflecting changes in electronic scattering and interfacial coupling at the Sb--Au interface.

Calculations for free-standing and epitaxial antimonene place Sb $p$-derived states in this energy range\cite{akturk2015single,lu2021observation,hu2023chiral}, while photoemission measurements and band-structure studies of Au(111) identify the underlying Au $5d$ manifold\cite{sheverdyaeva2016energy,smith1974photoemission}. Previous studies of Sb-derived surface phases on Au(111), including $\alpha$-antimonene on Au(111), further show that Sb-derived states can hybridize with Au-derived states at the interface, producing electronic states modified by Sb adsorption and surface reconstruction\cite{hu2023chiral,pierron2026surface,castro2025three}. This evidence suggests that the observed peak shifts and spectral-weight redistribution cannot be attributed to geometric folding alone, but instead reflect an interplay between geometric folding and Sb--Au orbital hybridization, in which hybridization further modifies the folded Au-derived states. 
Consistent with this interpretation, the EDC analysis reveals systematic peak shifts and linewidth modifications upon Sb adsorption, indicating that the deeper valence states are significantly modified as the surface reconstruction develops. Within this framework, the $(14\times14)$ phase can be viewed as an intermediate hybridization regime, whereas the Rec$(3\times\sqrt{3})$ phase corresponds to stronger Sb--Au coupling, for which an additional spectral component becomes experimentally resolved. These observations indicate that the Sb-induced reconstruction modifies the electronic structure over an energy range extending several electronvolts below $E_F$, demonstrating that Sb--Au orbital hybridization affects not only the near-$E_F$ states but also deeper valence bands.

\section*{Conclusion}

Through a systematic investigation of the coverage-dependent electronic structure of Sb/Au(111), we reveal the roles of reconstruction-induced scattering and Sb--Au hybridization in shaping the surface electronic states. While Umklapp scattering from the superlattice potential accounts for the geometric features of the Fermi surface in the Rec$(3\times\sqrt{3})$ phase, the pronounced reduction of the pocket size near the Fermi level indicates a substantial modification of the Au-derived $sp$ dispersion. This deviation from purely geometric band folding reflects the increasing importance of Sb--Au hybridization. Consistently, the evolution of deeper valence states exhibits clear modifications of the spectral line shape, evidencing hybridization between Sb $p$ and Au $d$ orbitals over a wide energy range.

These results show that the electronic structure of reconstructed noble-metal surfaces cannot, in general, be described by simple band folding of substrate states alone, but is instead governed by the interplay between superlattice-induced scattering and interfacial orbital hybridization. Our findings further suggest that Sb/Au(111) provides a suitable model system for understanding the relationship between geometric band folding and intrinsic electronic renormalization, and demonstrate that adsorbate--substrate hybridization offers an effective route for tuning surface electronic structure in low-dimensional systems, providing opportunities for designing functional interfaces with tailored electronic properties.

\section*{Methods}

\subsection*{Sample preparation}

All samples were prepared under ultra-high vacuum conditions with a base pressure below $2\times10^{-10}$ mbar. The Au(111) single crystal was cleaned by repeated cycles of Ar$^{+}$ ion sputtering and annealing at $600\,^\circ\mathrm{C}$ until sharp diffraction spots of the Au herringbone reconstruction were observed by LEED. Sb was deposited from a Knudsen cell operated at $420\,^\circ\mathrm{C}$ while the crystal itself was held at room temperature. The deposition rate of approximately $\SI{5.1}{\angstrom\per\hour}$ was determined via a quartz-crystal microbalance. Two representative surface phases were obtained at calibrated Sb coverages: a $(14\times14)$ phase at 0.25 ML and a Rec$(3\times\sqrt{3})$ phase at 0.7 ML. After deposition, the samples were annealed at $260\,^\circ\mathrm{C}$ for $\SI{30}{\minute}$ to enhance periodic ordering and phase formation. The resulting surface structures were characterized by low-energy electron diffraction (LEED). The surface periodicities were determined from LEED patterns and simulated using LEEDpat\cite{Hermann_LEEDpat}.

\subsection*{ARPES measurements}

ARPES measurements were performed in an ultra-high-vacuum chamber with a base pressure below $2\times10^{-10}$ mbar. Samples were transferred under vacuum using a UHV suitcase with a base pressure below $1\times10^{-9}$ mbar and subsequently degassed in situ. The sample phases were identified based on LEED characterization performed in the growth chamber prior to transfer. Photoelectrons were excited using a SPECS UVS 300 helium discharge lamp operated in non-monochromatized He I mode ($h\nu = 21.22$ eV) and analyzed with a SPECS Phoibos 150 hemispherical electron analyzer. The overall energy and angular resolutions were better than 8 meV and $0.3^\circ$, respectively. All measurements were carried out at room temperature.

\bibliography{sample}

\begin{thebibliography}{10}
\urlstyle{rm}
\expandafter\ifx\csname url\endcsname\relax
  \def\url#1{\texttt{#1}}\fi
\expandafter\ifx\csname urlprefix\endcsname\relax\def\urlprefix{URL }\fi
\expandafter\ifx\csname doiprefix\endcsname\relax\def\doiprefix{DOI: }\fi
\providecommand{\bibinfo}[2]{#2}
\providecommand{\eprint}[2][]{\url{#2}}

\bibitem{titmuss1996reconstruction}
\bibinfo{author}{Titmuss, S.}, \bibinfo{author}{Wander, A.} \& \bibinfo{author}{King, D.}
\newblock \bibinfo{journal}{\bibinfo{title}{Reconstruction of clean and adsorbate-covered metal surfaces}}.
\newblock {\emph{\JournalTitle{Chemical Reviews}}} \textbf{\bibinfo{volume}{96}}, \bibinfo{pages}{1291--1306} (\bibinfo{year}{1996}).

\bibitem{yang2019atomically}
\bibinfo{author}{Yang, T.} \emph{et~al.}
\newblock \bibinfo{journal}{\bibinfo{title}{Atomically thin 2\uppercase{D} transition metal oxides: Structural reconstruction, interaction with substrates, and potential applications}}.
\newblock {\emph{\JournalTitle{Advanced Materials Interfaces}}} \textbf{\bibinfo{volume}{6}}, \bibinfo{pages}{1801160} (\bibinfo{year}{2019}).

\bibitem{chen1998scanning}
\bibinfo{author}{Chen, W.}, \bibinfo{author}{Madhavan, V.}, \bibinfo{author}{Jamneala, T.} \& \bibinfo{author}{Crommie, M.}
\newblock \bibinfo{journal}{\bibinfo{title}{Scanning tunneling microscopy observation of an electronic superlattice at the surface of clean gold}}.
\newblock {\emph{\JournalTitle{Physical Review Letters}}} \textbf{\bibinfo{volume}{80}}, \bibinfo{pages}{1469} (\bibinfo{year}{1998}).

\bibitem{zhuang2018band}
\bibinfo{author}{Zhuang, J.} \emph{et~al.}
\newblock \bibinfo{journal}{\bibinfo{title}{Band gap modulated by electronic superlattice in blue phosphorene}}.
\newblock {\emph{\JournalTitle{ACS Nano}}} \textbf{\bibinfo{volume}{12}}, \bibinfo{pages}{5059--5065} (\bibinfo{year}{2018}).

\bibitem{sohail2015umklapp}
\bibinfo{author}{Sohail, H.~M.} \& \bibinfo{author}{Uhrberg, R.}
\newblock \bibinfo{journal}{\bibinfo{title}{Umklapp induced surface band structure of \uppercase{A}g/\uppercase{G}e(111) 6$\times$6}}.
\newblock {\emph{\JournalTitle{Surface Science}}} \textbf{\bibinfo{volume}{635}}, \bibinfo{pages}{55--60} (\bibinfo{year}{2015}).

\bibitem{jauernik2018probing}
\bibinfo{author}{Jauernik, S.}, \bibinfo{author}{Hein, P.}, \bibinfo{author}{Gurgel, M.}, \bibinfo{author}{Falke, J.} \& \bibinfo{author}{Bauer, M.}
\newblock \bibinfo{journal}{\bibinfo{title}{Probing long-range structural order in \uppercase{S}n\uppercase{P}c/\uppercase{A}g(111) by umklapp process assisted low-energy angle-resolved photoelectron spectroscopy}}.
\newblock {\emph{\JournalTitle{Physical Review B}}} \textbf{\bibinfo{volume}{97}}, \bibinfo{pages}{125413} (\bibinfo{year}{2018}).

\bibitem{chiniwar2019substrate}
\bibinfo{author}{Chiniwar, S.} \emph{et~al.}
\newblock \bibinfo{journal}{\bibinfo{title}{Substrate-mediated umklapp scattering at the incommensurate interface of a monatomic alloy layer}}.
\newblock {\emph{\JournalTitle{Physical Review B}}} \textbf{\bibinfo{volume}{99}}, \bibinfo{pages}{155408} (\bibinfo{year}{2019}).

\bibitem{marchenko2012giant}
\bibinfo{author}{Marchenko, D.} \emph{et~al.}
\newblock \bibinfo{journal}{\bibinfo{title}{Giant rashba splitting in graphene due to hybridization with gold}}.
\newblock {\emph{\JournalTitle{Nat. Commun.}}} \textbf{\bibinfo{volume}{3}}, \bibinfo{pages}{1232} (\bibinfo{year}{2012}).

\bibitem{ast2007giant}
\bibinfo{author}{Ast, C.~R.} \emph{et~al.}
\newblock \bibinfo{journal}{\bibinfo{title}{Giant spin splitting through surface alloying}}.
\newblock {\emph{\JournalTitle{Physical Review Letters}}} \textbf{\bibinfo{volume}{98}}, \bibinfo{pages}{186807} (\bibinfo{year}{2007}).

\bibitem{sakamoto2019impact}
\bibinfo{author}{Sakamoto, M.} \emph{et~al.}
\newblock \bibinfo{journal}{\bibinfo{title}{Impact of orbital hybridization at molecule--metal interface on carrier dynamics}}.
\newblock {\emph{\JournalTitle{J. Phys. Chem. C}}} \textbf{\bibinfo{volume}{123}}, \bibinfo{pages}{25877--25882} (\bibinfo{year}{2019}).

\bibitem{patil2016arpes}
\bibinfo{author}{Patil, S.} \emph{et~al.}
\newblock \bibinfo{journal}{\bibinfo{title}{\uppercase{ARPES} view on surface and bulk hybridization phenomena in the antiferromagnetic \uppercase{K}ondo lattice \uppercase{C}e\uppercase{R}h$_2$\uppercase{S}i$_2$}}.
\newblock {\emph{\JournalTitle{Nature Communications}}} \textbf{\bibinfo{volume}{7}}, \bibinfo{pages}{11029} (\bibinfo{year}{2016}).

\bibitem{hu2023chiral}
\bibinfo{author}{Hu, J.}, \bibinfo{author}{Liu, L.}, \bibinfo{author}{Wang, X.}, \bibinfo{author}{Chen, Y.~P.} \& \bibinfo{author}{Wells, J.~W.}
\newblock \bibinfo{journal}{\bibinfo{title}{Chiral quantum well \uppercase{R}ashba splitting in \uppercase{S}b monolayer on \uppercase{A}u(111)}}.
\newblock {\emph{\JournalTitle{arXiv preprint arXiv:2308.06814}}}  (\bibinfo{year}{2023}).

\bibitem{cantero2021synthesis}
\bibinfo{author}{Cantero, E.~D.} \emph{et~al.}
\newblock \bibinfo{journal}{\bibinfo{title}{Synthesis and characterization of a pure 2\uppercase{D} antimony film on \uppercase{A}u(111)}}.
\newblock {\emph{\JournalTitle{J. Phys. Chem. C}}} \textbf{\bibinfo{volume}{125}}, \bibinfo{pages}{9273--9280} (\bibinfo{year}{2021}).

\bibitem{castro2025three}
\bibinfo{author}{Villalobos~Castro, J. d.~J.} \emph{et~al.}
\newblock \bibinfo{journal}{\bibinfo{title}{Three-dimensional deformations in single-layer $\alpha$ antimonene and interaction with a \uppercase{A}u(111) surface from first principles}}.
\newblock {\emph{\JournalTitle{arXiv preprint arXiv:2501.10180}}}  (\bibinfo{year}{2025}).

\bibitem{marsusi2018graphene}
\bibinfo{author}{Marsusi, F.}, \bibinfo{author}{Fedorov, I.} \& \bibinfo{author}{Gerivani, S.}
\newblock \bibinfo{journal}{\bibinfo{title}{Graphene-induced band gap renormalization in polythiophene: a many-body perturbation study}}.
\newblock {\emph{\JournalTitle{Journal of Physics: Condensed Matter}}} \textbf{\bibinfo{volume}{30}}, \bibinfo{pages}{035002} (\bibinfo{year}{2018}).

\bibitem{katoch2018giant}
\bibinfo{author}{Katoch, J.} \emph{et~al.}
\newblock \bibinfo{journal}{\bibinfo{title}{Giant spin-splitting and gap renormalization driven by trions in single-layer \uppercase{WS}$_2$/h-\uppercase{BN} heterostructures}}.
\newblock {\emph{\JournalTitle{Nature Physics}}} \textbf{\bibinfo{volume}{14}}, \bibinfo{pages}{355--359} (\bibinfo{year}{2018}).

\bibitem{garcia2011renormalization}
\bibinfo{author}{Garcia-Lastra, J.~M.} \& \bibinfo{author}{Thygesen, K.~S.}
\newblock \bibinfo{journal}{\bibinfo{title}{Renormalization of optical excitations in molecules near a metal surface}}.
\newblock {\emph{\JournalTitle{Physical Review Letters}}} \textbf{\bibinfo{volume}{106}}, \bibinfo{pages}{187402} (\bibinfo{year}{2011}).

\bibitem{watson2019probing}
\bibinfo{author}{Watson, M.~D.} \emph{et~al.}
\newblock \bibinfo{journal}{\bibinfo{title}{Probing the reconstructed \uppercase{F}ermi surface of antiferromagnetic \uppercase{B}a\uppercase{F}e$_2$\uppercase{A}s$_2$ in one domain}}.
\newblock {\emph{\JournalTitle{npj Quantum Materials}}} \textbf{\bibinfo{volume}{4}}, \bibinfo{pages}{36} (\bibinfo{year}{2019}).

\bibitem{brouet2025unfolding}
\bibinfo{author}{Brouet, V.}, \bibinfo{author}{Vedant, A.}, \bibinfo{author}{Bertran, F.}, \bibinfo{author}{Le~F{\`e}vre, P.} \& \bibinfo{author}{Rubel, O.}
\newblock \bibinfo{journal}{\bibinfo{title}{Unfolding the kagome lattice to improve understanding of \uppercase{ARPES} in \uppercase{C}o\uppercase{S}n}}.
\newblock {\emph{\JournalTitle{Physical Review B}}} \textbf{\bibinfo{volume}{112}}, \bibinfo{pages}{115135} (\bibinfo{year}{2025}).

\bibitem{zhou2019interfacial}
\bibinfo{author}{Zhou, D.} \emph{et~al.}
\newblock \bibinfo{journal}{\bibinfo{title}{Interfacial effects on the growth of atomically thin film: \uppercase{G}roup \uppercase{VA} elements on \uppercase{A}u(111)}}.
\newblock {\emph{\JournalTitle{Advanced Materials Interfaces}}} \textbf{\bibinfo{volume}{6}}, \bibinfo{pages}{1901050} (\bibinfo{year}{2019}).

\bibitem{khan2021novel}
\bibinfo{author}{Khan, K.} \emph{et~al.}
\newblock \bibinfo{journal}{\bibinfo{title}{Novel synthesis, properties and applications of emerging group \uppercase{VA} two-dimensional monoelemental materials (2\uppercase{D}-\uppercase{X}enes)}}.
\newblock {\emph{\JournalTitle{Materials Chemistry Frontiers}}} \textbf{\bibinfo{volume}{5}}, \bibinfo{pages}{6333--6391} (\bibinfo{year}{2021}).

\bibitem{zhang2017topologically}
\bibinfo{author}{Zhang, P.} \emph{et~al.}
\newblock \bibinfo{journal}{\bibinfo{title}{Topologically entangled \uppercase{R}ashba-split \uppercase{S}hockley states on the surface of grey arsenic}}.
\newblock {\emph{\JournalTitle{Physical Review Letters}}} \textbf{\bibinfo{volume}{118}}, \bibinfo{pages}{046802} (\bibinfo{year}{2017}).

\bibitem{lu2022realization}
\bibinfo{author}{Lu, Q.} \emph{et~al.}
\newblock \bibinfo{journal}{\bibinfo{title}{Realization of unpinned two-dimensional dirac states in antimony atomic layers}}.
\newblock {\emph{\JournalTitle{Nature Communications}}} \textbf{\bibinfo{volume}{13}}, \bibinfo{pages}{4603} (\bibinfo{year}{2022}).

\bibitem{lee2015two}
\bibinfo{author}{Lee, J.}, \bibinfo{author}{Tian, W.-C.}, \bibinfo{author}{Wang, W.-L.} \& \bibinfo{author}{Yao, D.-X.}
\newblock \bibinfo{journal}{\bibinfo{title}{Two-dimensional pnictogen honeycomb lattice: structure, on-site spin-orbit coupling and spin polarization}}.
\newblock {\emph{\JournalTitle{Sci. Rep.}}} \textbf{\bibinfo{volume}{5}}, \bibinfo{pages}{11512} (\bibinfo{year}{2015}).

\bibitem{lei2019anisotropic}
\bibinfo{author}{Lei, T.} \emph{et~al.}
\newblock \bibinfo{journal}{\bibinfo{title}{Anisotropic electronic structure of antimonene}}.
\newblock {\emph{\JournalTitle{Applied Physics Letters}}} \textbf{\bibinfo{volume}{115}} (\bibinfo{year}{2019}).

\bibitem{zhu2020kagome}
\bibinfo{author}{Zhu, J.}, \bibinfo{author}{He, C.}, \bibinfo{author}{Zhao, Y.-H.} \& \bibinfo{author}{Fu, B.}
\newblock \bibinfo{journal}{\bibinfo{title}{Kagome-like group-\uppercase{VA} monolayers with indirect--direct band gap transition and anisotropic mobility}}.
\newblock {\emph{\JournalTitle{Journal of Materials Chemistry C}}} \textbf{\bibinfo{volume}{8}}, \bibinfo{pages}{2732--2740} (\bibinfo{year}{2020}).

\bibitem{novoselov2004electric}
\bibinfo{author}{Novoselov, K.~S.} \emph{et~al.}
\newblock \bibinfo{journal}{\bibinfo{title}{Electric field effect in atomically thin carbon films}}.
\newblock {\emph{\JournalTitle{Science}}} \textbf{\bibinfo{volume}{306}}, \bibinfo{pages}{666--669} (\bibinfo{year}{2004}).

\bibitem{zhu2015epitaxial}
\bibinfo{author}{Zhu, F.-f.} \emph{et~al.}
\newblock \bibinfo{journal}{\bibinfo{title}{Epitaxial growth of two-dimensional stanene}}.
\newblock {\emph{\JournalTitle{Nature Materials}}} \textbf{\bibinfo{volume}{14}}, \bibinfo{pages}{1020--1025} (\bibinfo{year}{2015}).

\bibitem{maniraj2019case}
\bibinfo{author}{Maniraj, M.} \emph{et~al.}
\newblock \bibinfo{journal}{\bibinfo{title}{A case study for the formation of stanene on a metal surface}}.
\newblock {\emph{\JournalTitle{Communications Physics}}} \textbf{\bibinfo{volume}{2}}, \bibinfo{pages}{12} (\bibinfo{year}{2019}).

\bibitem{ji2016two}
\bibinfo{author}{Ji, J.} \emph{et~al.}
\newblock \bibinfo{journal}{\bibinfo{title}{Two-dimensional antimonene single crystals grown by van der \uppercase{W}aals epitaxy}}.
\newblock {\emph{\JournalTitle{Nature Communications}}} \textbf{\bibinfo{volume}{7}}, \bibinfo{pages}{13352} (\bibinfo{year}{2016}).

\bibitem{lu2021observation}
\bibinfo{author}{Lu, Q.} \emph{et~al.}
\newblock \bibinfo{journal}{\bibinfo{title}{Observation of symmetry-protected dirac states in nonsymmorphic $\alpha$-antimonene}}.
\newblock {\emph{\JournalTitle{Physical Review B}}} \textbf{\bibinfo{volume}{104}}, \bibinfo{pages}{L201105} (\bibinfo{year}{2021}).

\bibitem{lu2024realization}
\bibinfo{author}{Lu, Q.} \emph{et~al.}
\newblock \bibinfo{journal}{\bibinfo{title}{Realization of a two-dimensional \uppercase{W}eyl semimetal and topological \uppercase{F}ermi strings}}.
\newblock {\emph{\JournalTitle{Nat. Commun.}}} \textbf{\bibinfo{volume}{15}}, \bibinfo{pages}{6001} (\bibinfo{year}{2024}).

\bibitem{zan2019antimony}
\bibinfo{author}{Zan, L.}, \bibinfo{author}{Xing, D.}, \bibinfo{author}{Abd-El-Latif, A.~A.} \& \bibinfo{author}{Baltruschat, H.}
\newblock \bibinfo{journal}{\bibinfo{title}{Antimony deposition onto \uppercase{A}u(111) and insertion of \uppercase{M}g}}.
\newblock {\emph{\JournalTitle{Beilstein Journal of Nanotechnology}}} \textbf{\bibinfo{volume}{10}}, \bibinfo{pages}{2541--2552} (\bibinfo{year}{2019}).

\bibitem{ma1993adsorption}
\bibinfo{author}{Ma, P.} \& \bibinfo{author}{Slavin, A.}
\newblock \bibinfo{journal}{\bibinfo{title}{Adsorption of antimony on \uppercase{A}u(111) at room temperature}}.
\newblock {\emph{\JournalTitle{Journal of Vacuum Science \& Technology A: Vacuum, Surfaces, and Films}}} \textbf{\bibinfo{volume}{11}}, \bibinfo{pages}{2003--2007} (\bibinfo{year}{1993}).

\bibitem{hu2025unconventional}
\bibinfo{author}{Hu, J.}, \bibinfo{author}{Wang, X.}, \bibinfo{author}{{\AA}sland, A.~C.} \& \bibinfo{author}{Wells, J.~W.}
\newblock \bibinfo{journal}{\bibinfo{title}{Unconventional broadening of \uppercase{R}ashba spin splitting in a \uppercase{A}u$_2$\uppercase{S}b surface alloy with periodic structural defects}}.
\newblock {\emph{\JournalTitle{npj Quantum Materials}}} \textbf{\bibinfo{volume}{10}}, \bibinfo{pages}{5} (\bibinfo{year}{2025}).

\bibitem{shah2021atomic}
\bibinfo{author}{Shah, J.}, \bibinfo{author}{Wang, W.}, \bibinfo{author}{Sohail, H.~M.} \& \bibinfo{author}{Uhrberg, R.}
\newblock \bibinfo{journal}{\bibinfo{title}{Atomic and electronic structures of the \uppercase{A}u$_2$\uppercase{S}n surface alloy on au(111)}}.
\newblock {\emph{\JournalTitle{Physical Review B}}} \textbf{\bibinfo{volume}{104}}, \bibinfo{pages}{125408} (\bibinfo{year}{2021}).

\bibitem{hochhaus2025square}
\bibinfo{author}{Hochhaus, J.~A.} \emph{et~al.}
\newblock \bibinfo{journal}{\bibinfo{title}{First evidence of a square-like \uppercase{S}n lattice on the \uppercase{A}u$_2$\uppercase{S}n surface alloy on \uppercase{A}u(111)}}.
\newblock {\emph{\JournalTitle{Applied Surface Science}}} \textbf{\bibinfo{volume}{714}}, \bibinfo{pages}{164470} (\bibinfo{year}{2025}).

\bibitem{hochhaus2025structural}
\bibinfo{author}{Hochhaus, J.~A.} \emph{et~al.}
\newblock \bibinfo{journal}{\bibinfo{title}{Structural analysis of \uppercase{S}n on \uppercase{A}u(111) at low coverages: \uppercase{T}owards the surface alloy with alternating fcc and hcp domains}}.
\newblock {\emph{\JournalTitle{Sci. Rep.}}} \textbf{\bibinfo{volume}{15}}, \bibinfo{pages}{7953} (\bibinfo{year}{2025}).

\bibitem{pierron2026surface}
\bibinfo{author}{Pierron, T.} \emph{et~al.}
\newblock \bibinfo{journal}{\bibinfo{title}{Surface reconstruction-driven band folding and spin-orbit enhancement at the $\alpha$-antimonene/\uppercase{A}u(111) interface}}.
\newblock {\emph{\JournalTitle{arXiv preprint arXiv:2601.02922}}}  (\bibinfo{year}{2026}).

\bibitem{zhang2011precise}
\bibinfo{author}{Richard, P.}, \bibinfo{author}{Qian, T.}, \bibinfo{author}{Xu, Y.-M.}, \bibinfo{author}{Dai, X.} \& \bibinfo{author}{Ding, H.}
\newblock \bibinfo{journal}{\bibinfo{title}{A precise method for visualizing dispersive features in image plots}}.
\newblock {\emph{\JournalTitle{Rev. Sci. Instrum.}}} \textbf{\bibinfo{volume}{82}} (\bibinfo{year}{2011}).

\bibitem{lashell1996spin}
\bibinfo{author}{LaShell, S.}, \bibinfo{author}{McDougall, B.} \& \bibinfo{author}{Jensen, E.}
\newblock \bibinfo{journal}{\bibinfo{title}{Spin splitting of an \uppercase{A}u(111) surface state band observed with angle resolved photoelectron spectroscopy}}.
\newblock {\emph{\JournalTitle{Physical Review Letters}}} \textbf{\bibinfo{volume}{77}}, \bibinfo{pages}{3419} (\bibinfo{year}{1996}).

\bibitem{reinert2001direct}
\bibinfo{author}{Reinert, F.}, \bibinfo{author}{Nicolay, G.}, \bibinfo{author}{Schmidt, S.}, \bibinfo{author}{Ehm, D.} \& \bibinfo{author}{H{\"u}fner, S.}
\newblock \bibinfo{journal}{\bibinfo{title}{Direct measurements of the \uppercase{L}-gap surface states on the (111) face of noble metals by photoelectron spectroscopy}}.
\newblock {\emph{\JournalTitle{Physical Review B}}} \textbf{\bibinfo{volume}{63}}, \bibinfo{pages}{115415} (\bibinfo{year}{2001}).

\bibitem{tusche2015spin}
\bibinfo{author}{Tusche, C.}, \bibinfo{author}{Krasyuk, A.} \& \bibinfo{author}{Kirschner, J.}
\newblock \bibinfo{journal}{\bibinfo{title}{Spin resolved bandstructure imaging with a high resolution momentum microscope}}.
\newblock {\emph{\JournalTitle{Ultramicroscopy}}} \textbf{\bibinfo{volume}{159}}, \bibinfo{pages}{520--529} (\bibinfo{year}{2015}).

\bibitem{courths1986electronic}
\bibinfo{author}{Courths, R.}, \bibinfo{author}{Zimmer, H.-G.}, \bibinfo{author}{Goldmann, A.} \& \bibinfo{author}{Saalfeld, H.}
\newblock \bibinfo{journal}{\bibinfo{title}{Electronic structure of gold: An angle-resolved photoemission study along the $\lambda$ line}}.
\newblock {\emph{\JournalTitle{Physical Review B}}} \textbf{\bibinfo{volume}{34}}, \bibinfo{pages}{3577} (\bibinfo{year}{1986}).

\bibitem{sheverdyaeva2016energy}
\bibinfo{author}{Sheverdyaeva, P.} \emph{et~al.}
\newblock \bibinfo{journal}{\bibinfo{title}{Energy-momentum mapping of d-derived \uppercase{A}u(111) states in a thin film}}.
\newblock {\emph{\JournalTitle{Physical Review B}}} \textbf{\bibinfo{volume}{93}}, \bibinfo{pages}{035113} (\bibinfo{year}{2016}).

\bibitem{smith1974photoemission}
\bibinfo{author}{Smith, N.~V.}, \bibinfo{author}{Wertheim, G.}, \bibinfo{author}{H{\"u}fner, S.} \& \bibinfo{author}{Traum, M.~M.}
\newblock \bibinfo{journal}{\bibinfo{title}{Photoemission spectra and band structures of d-band metals. \uppercase{IV}. \uppercase{X}-ray photoemission spectra and densities of states in \uppercase{R}h, \uppercase{P}d, \uppercase{A}g, \uppercase{I}r, \uppercase{P}t, and \uppercase{A}u}}.
\newblock {\emph{\JournalTitle{Physical Review B}}} \textbf{\bibinfo{volume}{10}}, \bibinfo{pages}{3197} (\bibinfo{year}{1974}).

\bibitem{anderson1976chemisorption}
\bibinfo{author}{Anderson, J.} \& \bibinfo{author}{Lapeyre, G.}
\newblock \bibinfo{journal}{\bibinfo{title}{Chemisorption-induced surface umklapp processes in angle-resolved synchrotron photoemission from w (001)}}.
\newblock {\emph{\JournalTitle{Physical Review Letters}}} \textbf{\bibinfo{volume}{36}}, \bibinfo{pages}{376} (\bibinfo{year}{1976}).

\bibitem{shirley1972high}
\bibinfo{author}{Shirley, D.~A.}
\newblock \bibinfo{journal}{\bibinfo{title}{High-resolution \uppercase{X}-ray photoemission spectrum of the valence bands of gold}}.
\newblock {\emph{\JournalTitle{Physical Review B}}} \textbf{\bibinfo{volume}{5}}, \bibinfo{pages}{4709} (\bibinfo{year}{1972}).

\bibitem{damascelli2003angle}
\bibinfo{author}{Damascelli, A.}, \bibinfo{author}{Hussain, Z.} \& \bibinfo{author}{Shen, Z.-X.}
\newblock \bibinfo{journal}{\bibinfo{title}{Angle-resolved photoemission studies of the cuprate superconductors}}.
\newblock {\emph{\JournalTitle{Reviews of Modern Physics}}} \textbf{\bibinfo{volume}{75}}, \bibinfo{pages}{473} (\bibinfo{year}{2003}).

\bibitem{hufner2013photoelectron}
\bibinfo{author}{H{\"u}fner, S.}
\newblock \emph{\bibinfo{title}{Photoelectron Spectroscopy: Principles and Applications}} (\bibinfo{publisher}{Springer Science \& Business Media}, \bibinfo{year}{2013}).

\bibitem{akturk2015single}
\bibinfo{author}{Akt{\"u}rk, O.~{\"U}.}, \bibinfo{author}{{\"O}z{\c{c}}elik, V.~O.} \& \bibinfo{author}{Ciraci, S.}
\newblock \bibinfo{journal}{\bibinfo{title}{Single-layer crystalline phases of antimony: Antimonenes}}.
\newblock {\emph{\JournalTitle{Physical Review B}}} \textbf{\bibinfo{volume}{91}}, \bibinfo{pages}{235446} (\bibinfo{year}{2015}).

\bibitem{Hermann_LEEDpat}
\bibinfo{author}{Hermann, K.}
\newblock \bibinfo{title}{{LEEDpat download package}}, \doiprefix\url{10.17617/3.8AYKWU} (\bibinfo{year}{2022}).

\end{thebibliography}

\section*{Acknowledgements}

Z.Z. and C.Wa. acknowledge valuable discussions with D.M.Janas, M.Capra, D.Sharma and J.A.Hochhaus.

\section*{Funding}

The spin-ARPES system has been financed by the Deutsche Forschungsgemeinschaft (DFG) through the project INST 212/411-1 FUGG and by the “Ministerium für Kultur und Wissenschaft des Landes Nordrhein-Westfalen”.

\section*{Author contributions statement}

Z.Z. and C.Wa. conceived the project. Z.Z. performed the ARPES experiments and analyzed the ARPES data. C.Wa. and S.H. prepared the samples and carried out the structural characterization. M.C. and C.We. supervised the project and contributed to the interpretation of the results. Z.Z. wrote the manuscript with input from all authors. All authors discussed the results and commented on the manuscript.

\section*{Additional information}

\noindent\textbf{Competing interests}
The authors declare no competing interests.\vspace{2mm}

\noindent\textbf{Use of AI-assisted tools in manuscript preparation}
During the preparation of this manuscript, large language models were used solely for language editing to improve grammar and readability. All outputs were carefully checked and revised by the authors, who take full responsibility for the final manuscript. No generative AI tools were used in the scientific aspects of this work, including data acquisition, analysis, or interpretation.

\end{document}